\documentclass[conference]{IEEEtran}
\IEEEoverridecommandlockouts
\usepackage{cite}
\usepackage{amsmath,amssymb,amsfonts}
\usepackage{algorithmic}
\usepackage{graphicx}
\usepackage{textcomp}
\usepackage{xcolor}
\def\BibTeX{{\rm B\kern-.05em{\sc i\kern-.025em b}\kern-.08em
    T\kern-.1667em\lower.7ex\hbox{E}\kern-.125emX}}

\usepackage{caption}
\usepackage{subcaption}
\usepackage{stfloats}
\graphicspath{{./Figures/}}

\newtheorem{Conclusion}{Conclusion}
\newtheorem{Lemma}{Lemma}
\newtheorem{Theorem}{Theorem}

\DeclareMathAlphabet{\mathit}{OT1}{bch}{m}{it}

\renewcommand{\vec}[1]{{\bf #1}}
\newcommand{\src}{{\textnormal{S}}}
\newcommand{\des}{{\textnormal{D}}}
\newcommand{\ris}{{\textnormal{R}}}
\newcommand{\uav}{{\textnormal{U}}}

\newcommand{\AB}{\textnormal{AB}}

\newcommand{\nodeA}{\textnormal{A}}
\newcommand{\nodeB}{\textnormal{B}}
\newcommand{\PL}{\ell}
\newcommand{\los}{{\sf LOS}}
\newcommand{\nlos}{{\sf NLOS}}

\newcommand{\snr}{{\gamma}}

\newcommand{\rhs}{{\tt RHS}}

\newcommand{\atog}{\mathsf{A2G}}
\newcommand{\gtoa}{\mathsf{G2A}}

\begin{document}

\title{Statistical Characterization of RIS-assisted UAV Communications in Terrestrial and Non-Terrestrial Networks Under Channel Aging \\
}

\author{\IEEEauthorblockN{
        Thanh~Luan~Nguyen\IEEEauthorrefmark{1}, 
        Georges~Kaddoum\IEEEauthorrefmark{1}\IEEEauthorrefmark{2}, 
        Tri~Nhu~Do\IEEEauthorrefmark{3},
        and Zygmunt J. Haas\IEEEauthorrefmark{4}
        }
    \IEEEauthorrefmark{1} Department of Electrical Engineering, \'{E}cole de Technologie Sup\'{e}rieure (\'{E}TS), Montr\'{e}al, QC, Canada \\
    \IEEEauthorrefmark{2} Artificial Intelligence \& Cyber Systems Research Center, Lebanese American University \\
    \IEEEauthorrefmark{3} Electrical Engineering Department, Polytechnique Montr\'{e}al, Montr\'{e}al, QC, Canada \\
    \IEEEauthorrefmark{4} Department of Computer Science, University of Texas at Dallas, TX 75080, USA \\ 
    \IEEEauthorrefmark{4} School of Electrical and Computer Engineering, Cornell University, Ithaca, NY 14853, USA
    \\
      Emails: 
        thanh-luan.nguyen.1@ens.etsmtl.ca,
        georges.kaddoum@etsmtl.ca, \\
        \IEEEauthorrefmark{2}tri-nhu.do@polymtl.ca,
        \IEEEauthorrefmark{3}haas@cornell.edu
        
}

\maketitle

\begin{abstract}
This paper studies the statistical characterization of ground-to-air (G2A) and reconfigurable intelligent surface (RIS)-assisted air-to-ground (A2G) communications with unmanned aerial vehicles (UAVs) in terrestrial and non-terrestrial networks under the impact of channel aging. 
    We first model the G2A and A2G signal-to-noise ratios (SNRs) as non-central complex Gaussian quadratic random variables (RVs) and derive their exact probability density functions, offering a unique characterization for the A2G SNR as the product of two scaled non-central chi-square RVs.
Moreover, we also find that, for a large number of RIS elements, the RIS-assisted A2G channel can be characterized as a single Rician fading channel. 
    Our results reveal the presence of channel hardening in A2G communication under low UAV speeds, where we derive the maximum target spectral efficiency (SE) for a system to maintain a consistent required outage level. 
Meanwhile, high UAV speeds, exceeding 50 m/s, lead to a significant performance degradation, which cannot be mitigated by increasing the number of RIS elements.
\end{abstract}

\begin{IEEEkeywords}
RIS, UAV, channel aging, channel characterization, outage probability
\end{IEEEkeywords}

\section{Introduction}
\IEEEPARstart{I}{ntegrated} terrestrial networks (TNs) and non-terrestrial networks (NTNs) are increasingly considered promising candidates for  sixth-generation (6G) wireless networks to provide ubiquitous and high-speed connectivity for various devices and applications \cite{GeraciCM2023, AzariCST2022}.
    TNs and NTNs consist of complex and diverse infrastructures, such as different types of ground base stations (BSs), stationary and mobile ground user equipments (GUEs).
However, these conventional infrastructures may face challenges in supporting demanding services, such as high-definition live streaming, remote communication, or emergency responses. 
    To overcome these challenges, three-dimensional (3D) flying unmanned aerial vehicle (UAV) platforms, such as Dà-Jiāng Innovations (DJI) drones, have been proposed as a promising solution \cite{FengIN2020,  MozaffariCM2021}.

In the context of TN-NTN systems, UAVs, characterized by their cost-effectiveness and adaptability, can offer ubiquitous coverage to support a diverse range of applications in the emerging 6G wireless networks \cite{FengIN2020}.
    Specifically, UAVs can be integrated with leading technologies that enable high-performance wireless communication even in remote and disaster-prone scenarios. 
    Some of these technologies are terahertz (THz) communication \cite{WangTVT2020}, massive multiple-input multiple-output (mMIMO) \cite{NguyenTCCN2022}, and reconfigurable intelligent surfaces (RIS), which can manipulate the propagation of electromagnetic waves and create favorable wireless channels \cite{YangTVT2020, BoulogeorgosTVT2022}.
    It is important to note that the mobility of UAVs introduces a unique challenge as it leads to variations in the Ground-to-Air (G2A) and Air-to-Ground (A2G) channels over time, a phenomenon termed {\it channel aging}. 
This phenomenon causes a mismatch between the current channel state information (CSI) and the estimated CSI, which can degrade the performance and reliability of UAV-integrated wireless systems \cite{ChopraTWC2018, MozaffariCST2019, ZhengTWC2021}.

A promising solution to combat channel aging is by integrating reconfigurable intelligent surfaces (RISs) \cite{ZhangTVT2023, JiangCL2023, PapazafeiropoulosTVT2023}.
    Using planar arrays of many low-cost passive elements, RISs can intelligently manipulate the direction of the impinging electromagnetic waves, thus improving coverage and reliability, especially when direct line-of-sight (LOS) links are sacred \cite{ChopraTWC2018, ZhengTWC2021, BansalTC2023}.
However, RIS operation relies on perfect channel state information (CSI) for optimal phase shift configuration, posing a challenge in dynamic communication environments like RIS-assisted UAV systems.
    While RIS-assisted UAV wireless communication has gained significant attention in the literature \cite{ChopraTWC2018, ZhengTWC2021, BansalTC2023}, there remains a theoretical gap concerning the true statistical characterization aspect of G2A and RIS-assisted A2G communication under channel aging. 
This research gap has resulted in a lack of insightful design parameters, such as how high can the spectral efficiency (SE) be without compromising the end-to-end (e2e) outage probability (OP) for different UAV speeds?

This paper aims to fill these gaps by studying the statistical characterization of G2A and RIS-assisted A2G communications under the impact of channel aging. 
    Specifically, we formulate the G2A SNR and the RIS-assisted A2G SNR as non-central complex Gaussian quadratic forms \cite{Mathai1992}. 
By exploiting Laplace/inverse Laplace transforms, we determine the exact probability density function (PDF) of the G2A SNR. 
    Then, we prove that the A2G SNR can be characterized as the product of two scaled non-central chi-square (SNCCS) RVs before determining its exact PDF. 
    The main contributions of this paper can be summarized as follows:

\begin{itemize}
    \item We provide exact and tractable PDFs for the G2A and A2G SNRs and provide various insights based on the derived analytical PDFs.
    \item We derive a tractable formula for the target SE for the systems to operate at various desired outage levels. 
    \item We show that the channel hardening phenomena can appear in A2G communication, primarily when the UAV speed is low.
    \item We demonstrate through numerical results that increasing the number of RIS elements and the BS's antennas can enhance the maximum target SE up to a certain limit. 
    \item We show that high UAV speeds lead to a substantial reduction in the target SE, which cannot be mitigated with additional RIS elements or antennas.
\end{itemize}

\section{System Model}

In this paper, we consider a UAV-aided wireless communication system, where a single-antenna GUE is served by an $M$-antenna BS with the help of a single-antenna UAV and a passive RIS with $N$ reflecting elements installed on a building facade to assist the A2G communications.  
    Hereafter, we use $\src$, $\src_m$, $\uav$, $\ris$, $\ris_n$, and $\des$ as acronyms for the BS, the BS's antenna $m \in {\cal M}$, the UAV, the RIS, the RIS's element $n \in {\cal N}$, and the ground UE, respectively, where ${\cal M} \triangleq \{ 1, 2, \dots, M \}$ and ${\cal N} \triangleq \{ 1, 2, \dots, N\}$.
We consider that the direct $\src$-$\des$, $\src$-$\ris$, and $\uav$-$\des$ links are unavailable due to various Radio Frequency (RF) obstructions. 
Moreover, an $\nodeA$-$\nodeB$ channel at time instant $t$ is modeled as ${\sqrt{\PL_\AB[t]} h_{\AB}[t]}$ with $\PL_\AB[n]$ representing the path loss component, and $h_{\AB}[t]$ representing the small-scale channel coefficient.
    For practical purposes, we consider the 3GPP UMi standard to model the {$\ris\text{-}\des$} channel's path loss \cite{3GPP}. Moreover, the 3GPP UMi-AV standard is adopted to model the path loss of the {$\src$-$\uav$} (G2A) and the {$\uav$-$\ris$} (A2G) channels \cite{3gpp20173gpp}.
Specifically, the 3GPP ${\text{UMi-AV}}$ standard imposes the LOS probability $p_\AB^\los[t]$ and the NLOS probability ${p_\AB^\nlos[t] = 1 - p_\AB^\los[t]}$ on the on a $\nodeA$-$\nodeB$ channels (i.e., G2A and A2G channels), which yields the corresponding path losses as ${\PL_\AB[t] = \PL_\AB^\los[t]}$ and ${\PL_\AB[t] = \PL_\AB^\nlos[t]}$ with probability $p_\AB^\los[t]$ and $p_\AB^\nlos[t]$, respectively. Nevertheless, our analysis can be applied to other scenarios with different path loss characteristics.
    In addition, we consider that all channels experience Rician-$K$ fading, where $h_{\AB}[t] = \frac{\sqrt{\kappa_\AB[t]} \check{h}_\AB^{\los}[t] + {h}^{\nlos}_\AB[t] }{\sqrt{\kappa_\AB[t] +1}}$ with $\check{h}_\AB^{\los}[t]$ denoting the deterministic LOS component, ${h}^{\nlos}_\AB[t]$ presenting the random scattering/NLOS component, modeled as a zero-mean complex Gaussian process with unit variance, and $\kappa_\AB[t]$ is the Rician-$K$ factor.
We consider that $\kappa_\AB[t] = K_0 e^{\frac{2 \theta_\AB[t]}{\pi}\operatorname{ln}\frac{K_\pi}{K_0}}$ under the LOS scenario, where $K_0$ and $K_\pi$ are environmental coefficient and $\theta_\AB[t]$ [rad] is the elevation angle at $\nodeB$ with respect to $\nodeA$. $\kappa_\AB[t] = 0$ in the NLOS scenario, indicating Rayleigh fading in rich scattering environments.

Due to the UAV's mobility, the propagation environment evolves over time which causes channel aging \cite{ZhangTVT2023, ZhengTWC2021, ChopraTWC2018}. 
    To measure the aging of CSI caused by the Doppler effect, correlation metrics are used to determine the time-varying property of the CSI.
Specifically, the channel state at time instant $t$ can be modeled as \cite{ZhengTWC2021, ChopraTWC2018}
\begin{align}
\vec{h}_{\AB}[t] 
    = \rho_\AB[t] \vec{h}_{\AB}[0] + \bar{\rho}_{\AB}[t] \vec{f}_{\AB}[t],
\label{eq:aging_model}
\end{align}
where 
    $[\vec{h}_{\src\uav}[t]]_m = h_{\src_m\uav}[t]$, $\forall m \in \mathcal{M}$,
    $[\vec{h}_{\uav\ris}[t]]_n = h_{\uav\ris_n}[t]$ and 
    $[\vec{h}_{\ris\des}[t]]_n = h_{\ris_n\des}[t]$, $\forall n \in \mathcal{N}$,
    $\bar{\rho}_{\AB}[t] = \sqrt{1-\rho_\AB^2[t]}$, 
    $\vec{h}_{\AB}[0]$ is the initial channel state, ${\vec{f}_{\AB}[t] \sim \mathcal{CN}({\bf 0}, \vec{I})}$ represents the independent innovation component at the $t$th time instance, and $\rho_{\AB}[t] \in [-1, 1]$ is the temporal correlation coefficient. 
    As in \cite{ChopraTWC2018, ZhengTWC2021}, we consider ${\rho_{\AB}[t] = J_0(2\pi f_d t T_s)}$, where $J_0(\cdot)$ is the zeroth-order Bessel function of the first kind, $T_s$ [sec] is the sampling period, and ${f_d = \frac{v f_c}{c}}$ [Hz] is the maximum Doppler shift, where $v$ [m/s] is the UAV's {\it instantaneous} speed, $f_c$ [Hz] is the carrier frequency, and $c$ [m/s] is the speed of light.
Although \eqref{eq:aging_model} is not the first order autoregressive model, where the current CSI is a function of its state at the previous time instance, within a coherence time block, the correlation coefficients can be adjusted to approximately match Jakes model \cite{ZhengTWC2021}.
    
Considering the channel estimate at time instant $\tau$ is perfect and is utilized for information processing at time instant $t$, where channel estimates at later time instants are proved to be inaccurate \cite{ZhengTWC2021}. 
    Then, the effective channel at time instant $t$ can be expressed in terms of the estimated channel at the $\tau$th time instant as \cite{ZhengTWC2021, PapazafeiropoulosTVT2023}
\begin{align}
\vec{c}_{\AB}[t] 
    =  \rho_{\AB}[\tau-t] {\vec{c}}_{\AB}[\tau] + \bar{\rho}_{\AB}[\tau-t] \vec{z}_{\AB}[t], 
\label{eq:hab_aging_estimate_a}
\end{align}
where $\vec{z}_{\AB}[t] \sim \mathcal{CN}(\mathbf{0}, \mathbf{I})$ denotes the independent innovation component that correlates $\vec{c}_{\AB}[t]$ and ${\vec{c}}_{\AB}[\tau]$.     
Hereafter, we drop the time indices and denote the delayed channel estimate as ${\widehat{\vec{c}}_{\AB} = {\vec{c}}_{\AB}[\tau]}$ for convenience.
\begin{Conclusion}
We observe that an increase in UAV speed or a higher sample index yields a non-monotonic decrease in the correlation coefficient which oscillates around zero with decreasing magnitude.
\end{Conclusion}

\subsection{Ground-to-Air Signal-to-Noise Ratio}

In the first time slot, $\src$ uses the beamforming vector $\vec{w}_{\src\uav}$ to steer the desired signal $x_\src$ to $\uav$ with $\mathbb{E}[|x_{\src}|^2] = 1$. Hence, the received signal at $\uav$ is given by
\begin{align} \label{Eq10}
y_{\uav} &= \sqrt{P_\src} 
    (\rho_{\src\uav} \widehat{\vec{h}}_{\src\uav} 
        + {\textstyle\sqrt{1-\rho_{\src\uav}^2}} \vec{z}_{\src\uav})^{\sf H} \vec{w}_{\src\uav} x_{\src}
    + n_\uav,
\end{align}
where $P_\src$ [W] is the transmit power at $\src$ and $n_{\uav}$ is the additive white Gaussian noise (AWGN) with zero mean and variance~$\sigma^2_{\uav}$ [W]. 
We consider that $\src$ adopts maximal ratio transmission (MRT) based on the delayed CSI, where ${\vec{w}_{\src\uav} = \frac{\widehat{\vec{h}}_{\src\uav}}{\Vert\widehat{\vec{h}}_{\src\uav}\Vert}}$. 
Hence, the G2A SNR $\snr_{\gtoa} = \bar{\gamma}_{\src\uav} \frac{| {\vec{h}}_{\src\uav}^{\sf H} \widehat{\vec{h}}_{\src\uav} |^2}{\Vert \widehat{\vec{h}}_{\src\uav} \Vert^2}$, where $\bar{\gamma}_{\src\uav} = \frac{P_\src\PL_{\src\uav}}{\sigma_\uav^2}$, is further rewritten as
\begin{align} \label{Eq11}
\snr_{\gtoa}
    = \bar{\gamma}_{\src\uav}
    \left| 
        (\rho_{\src\uav} \widehat{\vec{h}}_{\src\uav} + \bar{\rho}_{\src\uav} {\vec{z}}_{\src\uav} )^{\sf H} \widehat{\vec{h}}_{\src\uav}
    \right|^2 \left\Vert \widehat{\vec{h}}_{\src\uav} \right\Vert^{-2}.
\end{align}
\subsection{Air-to-Ground Signal-to-Noise Ratio}

In the second time slot, $\uav$ decodes $x_\src$ and forwards the re-encoded version ${\widehat{x}_{\uav}}$ to $\des$ through $\ris$.
    Next, the RIS  reflects the signal $\widehat{x}_{\uav}$ from $\uav$ to $\des$ through $\ris$ by intelligently adjusting its phase-shift matrix.
We denote ${ \vec{\Theta} = \text{diag}([\beta_1 e^{j\vartheta_1}, \beta_2 e^{j\vartheta_2},\dots ,\beta_N e^{j\vartheta_N}]) }$ as the diagonal phase-shift matrix of the RIS, where $\beta_n \in \mathbb{C}$ denotes the amplitude reflection coefficient for each element, and ${ \vartheta_n \!=\! [0,2\pi) }$ [rad] denotes the phase shift of reflecting element ${ n \in {\cal N} }$ \cite{ZhangTVT2023, JiangCL2023}. For simplicity we assume that $\beta_1 = \beta_2 = \cdots = \beta_N = 1$ as indicated in \cite{ZhangTVT2023, JiangCL2023, PapazafeiropoulosTVT2023, BansalTC2023}.
Hence, the received signal at $\des$ is obtained~as
\begin{align}\label{Eq15}
y_{\des}
    =  \sqrt{P_{\uav} {\PL}_{\uav\ris} {\PL}_{\ris\des}} 
    \vec{h}_{\ris\des}^{\sf T} {\bf \Theta} {\vec{h}}_{\uav\ris} \widehat{x}_\uav + n_\des,
\end{align}
where $P_{\uav}$ [W] is the UAV's transmit power and $n_\des$ denotes the zero mean and variance $\sigma_\uav^2$ [W] AWGN at $\des$. Hence, the received A2G SNR is obtained as~\cite{JiangCL2023}
\begin{align}\label{Eq16}
\snr_\atog({\boldsymbol{\Theta}})
    &=  \frac{P_\uav \ell_{\uav\ris} \ell_{\ris\des}}{\sigma_\des^2}
    \left| 
        \vec{h}_{\ris_n\des}^{\sf T} {\bf \Theta} (\rho_{\uav\ris} \widehat{\vec{h}}_{\uav\ris} + \bar{\rho}_{\uav\ris} \vec{z}_{\uav\ris})
    \right|^2.
\end{align}

In practice, the number of phase-shifts is limited and constrained by the phase-shift resolution, denoted by ${Q \triangleq 2^b}$, where $b$ [bit] is the number of quantization bits and the value of a phase shift belongs to the set ${\cal Q} = \left\{0, \frac{2\pi}{Q}, \frac{4\pi}{Q},...,\frac{2\pi(Q-1)}{Q} \right\}$ \cite{Huang_TWC_2019}.
    Theoretically, with high phase-shift resolution, the RIS can be intelligently configured as $\vartheta_n = -\phi_{\ris_n\des} - \phi_{\uav\ris_n}$, $\forall n \in {\cal N}$, and can eliminate the phase error, where $\phi_{\ris_n\des}$ [rad] and $\phi_{\uav\ris_n}$ [rad] are the phase of $h_{\ris_n\des}$ and $h_{\uav\ris_n}$, respectively.
However, due to the delayed CSI, the true value of $\phi_{\uav\ris_n}$ is unknown, and only an approximate estimate $\widehat{\phi}_{\uav\ris_n} = \angle \widehat{h}_{\uav\ris_n}$ is available.
    This leads to an imperfect phase-shift configuration, given by $\widehat{\vartheta}_n = - \phi_{\ris_n\des} - \widehat{\phi}_{\uav\ris_n}$, $\forall n \in {\cal N}$.

\section{Statistical Characterization and Performance Analysis}

In this section, we first determine the statistical characteristics of the channel in the considered UAV-RIS system. 
Based on \cite{MozaffariCST2019}, the PDF of the G2A/A2G SNR can be modeled as a mixture of two distributions: the distribution in the LOS and NLOS scenarios. Specifically, the PDF of $\snr_{\gtoa}$ and $\snr_{\atog}$ are formulated as
\begin{align}
f_{\snr_{\gtoa}}(x) &= p^{\los}_{\src\uav} f^{\los}_{\snr_{\gtoa}}(x)
    + p^{\nlos}_{\src\uav} f^{\nlos}_{\snr_{\gtoa}}(x),~x > 0, \label{eq:pdf_gtoa_genForm} \\
f_{\snr_{\atog}}(x) &= p^{\los}_{\uav\ris} f^{\los}_{\snr_{\atog}}(x)
    + p^{\nlos}_{\uav\ris} f^{\nlos}_{\snr_{\atog}}(x),~x > 0, \label{eq:pdf_atog_genForm} 
\end{align}
respectively, where the explicit formulas for $p^{\los}_{\src\uav}$, $p^{\nlos}_{\src\uav}$, $p^{\los}_{\uav\ris}$, $p^{\nlos}_{\uav\ris}$ are presented in \cite{3GPP},
$f^{\los}_{\snr_{\gtoa}}(x)$, $f^{\los}_{\snr_{\atog}}(x)$ and $f^{\nlos}_{\snr_{\gtoa}}(x)$, $f^{\nlos}_{\snr_{\atog}}(x)$ are the PDFs of the G2A and A2G SNRs in the LOS and NLOS scenarios, respectively.
    Hereafter, we will focus on deriving $f^{\los}_{\snr_{\gtoa}}(x)$ and $f^{\los}_{\snr_{\atog}}(x)$, which are more challenging than deriving $f^{\nlos}_{\snr_{\gtoa}}(x)$ and $f^{\los}_{\snr_{\gtoa}}(x)$. 
It is noted that $f^{\nlos}_{\snr_{\gtoa}}(x)$ and $f^{\nlos}_{\snr_{\atog}}(x)$ are special cases of $f^{\los}_{\snr_{\gtoa}}(x)$ and $f^{\los}_{\snr_{\atog}}(x)$, respectively, with the Rician-$K$ factor being~zero.

\subsection{Statistical Characterization of G2A communication}
\begin{Theorem} \label{theo:pdf_cdf_g2asnr}
Under the LOS scenario, the exact PDF of the G2A SNR is formulated as
\begin{align}
f^{\los}_{\snr_{\gtoa}}(x)
    &=  e^{ -M \frac{\kappa_{\src\uav} \rho_{\src\uav}^2}{\kappa_{\src\uav} \bar{\rho}_{\src\uav}^2+1}
    -\frac{(\kappa_{\src\uav}+1)x}{\bar{\snr}_{\src\uav} (\kappa_{\src\uav} \bar{\rho}_{\src\uav}^2 + 1) } }
    \left(
        \frac{\kappa_{\src\uav}+1}{\kappa_{\src\uav} \bar{\rho}_{\src\uav}^2+1}
    \right)^M
    \nonumber\\
    &\quad\times
    \frac{1}{\bar{\snr}_{\src\uav}^{M}}
    \sum_{m=0}^{M-1}
    \binom{M-1}{m}
    (\bar{\snr}_{\src\uav} \bar{\rho}_{\src\uav}^2)^{m} ({\rho}_{\src\uav}^2)^{M-m-1}
    \nonumber\\
    &\quad\times
    \left( \frac{x}{\Xi_{\uav}} \right)^{\frac{M-m-1}{2}}
    I_{M-m-1} \left(
        \frac{ 2 \sqrt{\Xi_{\uav} x} }{\kappa_{\src\uav} \bar{\rho}_{\src\uav}^2+1}
    \right),
\label{eq:PDF_G2ASNR_LOS_F}
\end{align}
where $\Xi_{\uav} \triangleq M \kappa_{\src\uav} \rho_{\src\uav}^2 (\kappa_{\src\uav} +1) \bar{\snr}_{\src\uav}^{-1}$.
\end{Theorem}

\begin{IEEEproof}
Due to the complexity and length of the proof, we only provide here the key steps in the following. 
First, we rewrite the G2A SNR obtained in \eqref{Eq11} in the complex Gaussian quadratic form as $\snr_{\gtoa} = \snr_\uav^\ast \bar{\gamma}_{\src\uav} \snr_\uav$, where $\snr_\uav = \rho_{\src\uav} \Vert \widehat{\vec{h}}_{\src\uav} \Vert + \bar{\rho}_{\src\uav} \frac{ {\vec{z}}_{\src\uav}^{{\sf H}} \widehat{\vec{h}}_{\src\uav} }{ \Vert \widehat{\vec{h}}_{\src\uav} \Vert }$.
Here, ${\frac{ {\vec{z}}_{\src\uav}^{{\sf H}} \widehat{\vec{h}}_{\src\uav} }{ \Vert \widehat{\vec{h}}_{\src\uav} \Vert }}$ is a zero-mean complex Gaussian RV with unit variance which is independent of $ \Vert \widehat{\vec{h}}_{\src\uav} \Vert $, thus ${\snr_\uav \sim \mathcal{CN}(\rho_{\src\uav} \Vert \widehat{\vec{h}}_{\src\uav} \Vert, \bar{\rho}_{\src\uav}^2)}$. 
Hence, the Laplace Transform of $\snr_{\gtoa}$ conditioned on $\widehat{\vec{h}}_{\src\uav}$ is obtained as \cite{Mathai1992}
\begin{align}
{\cal L}_{\snr_{\gtoa}}(s\vert \widehat{\vec{h}}_{\src\uav})
    &=    e^{- \frac{ \bar{\snr}_{\src\uav} \rho_{\src\uav}^2 s }{ 1+s \bar{\gamma}_{\src\uav} \bar{\rho}_{\src\uav}^2 } \Vert \widehat{\vec{h}}_{\src\uav} \Vert^2 } 
    [ 1+s \bar{\snr}_{\src\uav} \bar{\rho}_{\src\uav}^2 ]^{-1}.
    \label{eq:LT_g2asnr_1}
\end{align}

It is highlighted that $\Vert \widehat{\vec{h}}_{\src\uav} \Vert^2$ is a SNCCS-distributed RV with scale factor $\frac{1}{2(\kappa_{\src\uav}+1)}$, $2M$ degree of freedom, non-centrality parameter $2M\kappa_{\src\uav}$. 
    In general, for $x > 0$, the PDF and CDF of a SNCCS-distributed RV with scale factor $\frac{\Omega}{2}$, $2 k$ degrees of freedom, and non-centrality parameter $2 \lambda$ are given by \cite{Mathai1992}
\begin{align}
f_{\chi_{2k}^2}(\Omega, \lambda ; x)
&=   \frac{e^{ -\frac{\chi}{\Omega} - \lambda }}{\Omega^{\frac{k+1}{2}}}
    \bigg(
        \frac{\chi}{\lambda}
    \bigg)^{\frac{k-1}{2}}
    I_{k-1}
    \bigg(
        2 \sqrt{\frac{\chi}{\Omega} \lambda}
    \bigg), \label{eq:pdf_snccs} \\
F_{\chi_{2k}^2}(\Omega, \lambda ; x)
&=  1 - Q_{k}\left(
        \sqrt{2} \sqrt{\lambda}, \sqrt{2} \sqrt{x / \Omega}
    \right), \label{eq:cdf_snccs}
\end{align}
respectively.
The Laplace transform of $\snr_\gtoa$ is obtained by averaging ${\cal L}_{\snr_{\gtoa}}(s\vert \widehat{\vec{h}}_{\src\uav})$ over $\Vert \widehat{\vec{h}}_{\src\uav} \Vert^2$, which is obtained as
\begin{align}
{\cal L}_{\snr_{\gtoa}}^{\los}(s)
    =   e^{-M\kappa_{\src\uav}}
    \frac{(1+\Delta_1 s)^{M-1}}{(1+\Delta_2 s)^M}
    e^{M\kappa_{\src\uav} \frac{1+\Delta_1 s}{1+\Delta_2 s} },
\label{eq:laplace_1}
\end{align}
where ${\Delta_1 = \bar{\gamma}_{\src\uav}\bar{\rho}_{\src\uav}^2}$,
    ${\Delta_2 = \bar{\gamma}_{\src\uav} \frac{\kappa_{\src\uav} \bar{\rho}_{\src\uav}^2+1}{\kappa_{\src\uav}+1}}$. 
Then, the PDF of $\snr_\gtoa$ is the inverse Laplace transform of ${\cal L}_{\snr_{\gtoa}}(s)$ from $s$-domain to $x$-domain. Hence,
\begin{align}
f_{\snr_{\gtoa}}^{\los}(x)
    &= e^{-M\kappa_{\src\uav}}
    e^{-\frac{x}{\Delta_2}}
    (\Delta_1)^{M-1} (\Delta_2)^{-M}
    e^{M \kappa_{\src\uav} \frac{\Delta_1}{\Delta_2}}
    e^{-\frac{x}{\Delta_3}}
    \nonumber\\
    &\quad\times
    \mathcal{L}^{-1}
    \Bigg\{
        \frac{s^{M-1}}{(s-\Delta_3^{-1})^M}
        e^{ M\kappa_{\src\uav} 
        \frac{\frac{\Delta_1}{\Delta_2 \Delta_3 }}{s-\Delta_3^{-1}} };
        s, x
    \Bigg\}, 
    \label{eq_pdf_GammaSU_2} \\
    &= e^{-M\kappa_{\src\uav}}
    e^{-\frac{x}{\Delta_2}}
    (\Delta_1)^{M-1} (\Delta_2)^{-M}
    e^{M \frac{\kappa_{\src\uav} \Delta_1}{\Delta_2}}
    e^{-\frac{x}{\Delta_3}}
    \nonumber\\
    &\quad\times 
    \frac{{\rm d}^{M-1}}{{\rm d} x^{M-1}} 
    e^{\frac{x}{\Delta_3}}
    \mathcal{L}^{-1}
    \Bigg\{
        \frac{e^{ M\kappa_{\src\uav} 
        \frac{\Delta_1}{\Delta_2 \Delta_3 }
        \frac{1}{s} }}{s^M};
        s, x
    \Bigg\},
    \label{eq_pdf_GammaSU_4}
\end{align}
where ${\Delta_3 \triangleq \frac{1}{\frac{1}{\Delta_1}-\frac{1}{\Delta_2}} = \frac{\Delta_1 \Delta_2}{\Delta_2-\Delta_1}}$. Here, the frequency shifting property and the derivative in the $x$-domain property of the Laplace transform are used to obtain \eqref{eq_pdf_GammaSU_2} and \eqref{eq_pdf_GammaSU_4}, respectively. 
    Using the Leibnitz's rule for higher order derivatives and \cite[Eq. (1.13.1.5)]{brychkov2008handbook}, the derivative part in \eqref{eq_pdf_GammaSU_4} is derived~as
\begin{align}
\Xi(x) 
&= \frac{{\rm d}^{M-1}}{{\rm d} x^{M-1}} 
e^{\frac{x}{\Delta_3}} x^{\frac{M-1}{2}}
\frac{I_{M-1}\left( 2 \sqrt{M \kappa_{\src\uav} 
    \frac{\Delta_1}{\Delta_2 \Delta_3} x} \right)}{ \left(M \kappa_{\src\uav} 
    \Delta_1 \Delta_2^{-1} \Delta_3^{-1}\right)^{\frac{M-1}{2}} }
\label{eq_pdf_GammaSU_5} \\
&= \sum_{m=0}^{M-1} \binom{M-1}{m} 
\frac{ e^{\frac{x}{\Delta_3}} x^{\frac{M-m-1}{2}} }{ \left( M \kappa_{\src\uav} \Delta_1 \Delta_2^{-1} \Delta_3^{-1} \right)^{\frac{M-m-1}{2}} }
\nonumber\\
&\qquad\qquad\times
I_{M-m-1}\left(
    2 \sqrt{M \kappa_{\src\uav} \Delta_1 \Delta_3 \Delta_2^{-1} x}
\right).
\label{eq_pdf_GammaSU_6}
\end{align}
Substituting \eqref{eq_pdf_GammaSU_6} into \eqref{eq_pdf_GammaSU_4} and after some mathematical manipulations, we obtain \eqref{eq:PDF_A2GSNR_LOS_F}. 
\end{IEEEproof}

Based on \eqref{eq:PDF_G2ASNR_LOS_F}, we obtain the following results:

i) The PDF of $\snr_\gtoa$ in the NLOS scenario is
\begin{IEEEeqnarray}{rCl}
    f_{\snr_{\gtoa}}^{\nlos}(x)
    &=   \frac{1}{\bar{\gamma}_{\src\uav}^M}
    \sum_{m=0}^{M-1} \binom{M-1}{m}
    \frac{ (\rho_{\src\uav}^2)^{M-m-1} }{(M-m-1)!} 
    \nonumber\\
    &\quad\times
     (\bar{\gamma}_{\src\uav} \bar{\rho}_{\src\uav}^2)^m x^{M-m-1} e^{-\frac{x}{\bar{\gamma}_{\src\uav}}},
\label{eq:pdf_g2asnr_nlos_f}
\end{IEEEeqnarray}
which reduces to $e^{-\frac{x}{\bar{\snr}_{\src\uav}}}$ when $\rho_{\src\uav} = 0$.

ii) In the high SNR regime, \eqref{eq:PDF_G2ASNR_LOS_F} is simplified as
\begin{IEEEeqnarray}{rCl}
f^{\los}_{\snr_\gtoa}(x)
    &\to& 
    \frac{e^{ -M \frac{\kappa_{\src\uav} \rho_{\src\uav}^2}{\kappa_{\src\uav} \bar{\rho}_{\src\uav}^2+1} }}{\bar{\snr}_{\src\uav}}
    \frac{(\kappa_{\src\uav}+1)^M (\bar{\rho}_{\src\uav}^2)^{M-1}}{(\kappa_{\src\uav} \bar{\rho}_{\src\uav}^2+1)^M},
\label{eq:pdf_snrG2A_los}
\end{IEEEeqnarray}
and \eqref{eq:pdf_g2asnr_nlos_f} is simplified as
\begin{IEEEeqnarray}{rCl}
f^{\nlos}_{\snr_\gtoa}(x)
    &\to& (\bar{\rho}_{\src\uav}^2)^{M-1}  \bar{\snr}_{\src\uav}^{-1}.
\label{eq:pdf_snrG2A_nlos}
\end{IEEEeqnarray}

iii) The CDF of $\snr_\gtoa$, denoted as $F_{\snr_\gtoa}(x)$, is upper bounded by its asymptotic formula. 
Based on \eqref{eq:pdf_gtoa_genForm}, \eqref{eq:pdf_snrG2A_los} and \eqref{eq:pdf_snrG2A_nlos} and since $F_{\snr_\gtoa}(x) = \int^x_0 f_{\snr_\gtoa}(x) {\rm d} x$ by definition, we obtain
\begin{align}
    F_{\snr_\gtoa}(x) \le p^{\los}_{\src\uav} f^{\los}_{\snr_\gtoa}(x) x + p^{\nlos}_{\src\uav} f^{\nlos}_{\snr_\gtoa}(x) x.
    \label{eq:cdf_G2ASNR_bound}
\end{align}

\subsection{Statistical Characterization of A2G communication}

Under imperfect phase shift, the A2G SNR can be expressed in a complex Gaussian quadratic form as $\snr_{\atog} = \snr_\ris^\ast \bar{\snr}_{\atog} \snr_\ris$, with
$\snr_\ris
    =  \sum_{n=1}^{N} g_{\ris_n\des}
    \big(
        \rho_{\uav\ris} \widehat{g}_{\uav\ris_n}
        + \bar{\rho}_{\uav\ris} z_{\uav\ris_n} e^{-j \hat{\phi}_{\uav\ris_n}}
    \big)$,
where $g_{\ris_n\des} = \left| h_{\ris_n\des} \right|$ and $\widehat{g}_{\uav\ris_n} = \left| \widehat{h}_{\uav\ris_n} \right|$.

\begin{Lemma}
\label{lem:1}
Under the imperfect phase-shift configuration, the A2G SNR is characterized by the product of two independent SNCCS-distributed RVs.
Specifically, we have
$\snr_\atog \sim \bar{\snr}_\atog  \bar{\rho}_{\uav\ris}^2 \Vert \vec{h}_{\ris\des} \Vert^2 |\chi|^2$ with $\chi = \mathcal{CN}\left( 
    \frac{\rho_{\uav\ris}}{\bar{\rho}_{\uav\ris}} 
    \frac{{\vec{g}}_{\ris\des}^{\sf T} \widehat{\vec{g}}_{\uav\ris}}{\Vert \vec{g}_{\ris\des} \Vert}, 1
\right)$.
\end{Lemma}
\begin{IEEEproof}
Let $\sum_{n=1}^{N} g_{\ris_n\des} \widehat{g}_{\uav\ris_n} =  \vec{g}_{\ris\des}^{\sf T} \widehat{\vec{g}}_{\uav\ris}$, where $\vec{g}_{\ris\des} \triangleq [|h_{\ris_1\des}|, \dots, |h_{\ris_N\des}|]^{\sf T}$ and $\widehat{\vec{g}}_{\uav\ris} \triangleq \left[\left| \widehat{h}_{\uav_1\ris} \right|, \dots, \left| \widehat{h}_{\uav_N\ris} \right| \right]^{\sf T}$, thus 
    $\snr_\ris
    \sim 
    \mathcal{CN}\left( 
        \rho_{\uav\ris} {\vec{g}}_{\ris\des}^{\sf T} \widehat{\vec{g}}_{\uav\ris}, 
        \bar{\rho}_{\uav\ris}^2 \Vert \vec{g}_{\ris\des} \Vert^2
    \right)$. In other words, we have
\begin{align}
\left|\snr_\ris\right|^2
    \sim 
    \bar{\rho}_{\uav\ris}^2 \Vert \vec{g}_{\ris\des} \Vert^2 
    \Bigg|
        \mathcal{CN}\Bigg( 
            \frac{\rho_{\uav\ris}}{\bar{\rho}_{\uav\ris}} 
            \frac{{\vec{g}}_{\ris\des}^{\sf T} \widehat{\vec{g}}_{\uav\ris}}{\Vert \vec{g}_{\ris\des} \Vert}, 1
        \Bigg)
    \Bigg|^2.
\end{align}

We note that $\frac{{\vec{g}}_{\ris\des}^{\sf T} \widehat{\vec{g}}_{\uav\ris}}{\Vert \vec{g}_{\ris\des} \Vert \Vert \widehat{\vec{g}}_{\uav\ris} \Vert}$ is the cosine similarity between two channel-gain vectors, $\vec{g}_{\ris\des}$ and $\widehat{\vec{g}}_{\uav\ris}$, which is independent of their corresponding vector length, i.e., $\Vert \vec{g}_{\ris\des} \Vert$ and $ \Vert \widehat{\vec{g}}_{\uav\ris} \Vert$.
    Hence $\Vert \vec{g}_{\ris\des} \Vert$ and $\frac{{\vec{g}}_{\ris\des}^{\sf T} \widehat{\vec{g}}_{\uav\ris}}{\Vert \vec{g}_{\ris\des} \Vert}$ are also independent RVs, and thus $\chi_\ris$ is independent of $\Vert \vec{g}_{\ris\des} \Vert^2$.
Then, the quadratic form $\chi^\ast \chi = |\chi|^2$ follows a SNCCS distribution. 
    Since $\Vert \vec{g}_{\ris\des} \Vert^2$ also follows a SNCCS distribution, the A2G SNR is the product of two SNCCS-distributed RVs. This completes the proof of Lemma~\ref{lem:1}.
\end{IEEEproof}

\begin{Theorem}
\label{theo:2}
Under the imperfect phase-shift configuration, the exact PDF of the A2G SNR in LOS scenarios is obtained~as
\begin{IEEEeqnarray}{rCl}
f_{\snr_\atog}^{\los}(x)
    &=&  
    \sum_{m=0}^{\infty}
    \sum_{j=0}^{m}
        \frac{ (N\kappa_{\ris\des})^j (\lambda_\ris)^{m-j} e^{-N\kappa_{\ris\des}-\lambda_\ris} }{j! (m-j)!} 
        \frac{\Xi_\ris^{\frac{N+k_\ris+m}{2}}}{2^{N+k_\ris+m-1}}
    \nonumber\\
    &&\quad\times
    \frac{K_{N-k_\ris-m+2j}\left( \sqrt{\Xi_\ris x} \right)}{(N+j-1)! \Gamma(m-j+k_\ris)}
    x^{\frac{N+k_\ris+m}{2}-1}.
\label{eq:PDF_A2GSNR_LOS_F}
\end{IEEEeqnarray}
where $\Xi_\ris \triangleq 4 (k_{\ris\des}+1) (\bar{\rho}_{\uav\ris}^2 \bar{\snr}_\atog \Omega_\ris)^{-1}$.
\end{Theorem}
\begin{IEEEproof}
It is noted that the mean and variance of $\chi$ are
\begin{align}
\mu_{\chi}
    &=  \frac{\rho_{\uav\ris}}{\bar{\rho}_{\uav\ris}} 
    \mathbb{E}\bigg\{
        \frac{{\vec{g}}_{\ris\des}^{\sf T}}{\Vert \vec{g}_{\ris\des} \Vert}
        \mathbb{E}\{ \widehat{\vec{g}}_{\uav\ris} \}
    \bigg\}, 
\label{eq:mean_chi_R} \\
\sigma_{\chi}^2
    &=  1 + \frac{\rho_{\uav\ris}^2}{\bar{\rho}_{\uav\ris}^2} 
    \mathbb{E}\bigg\{
        \frac{{\vec{g}}_{\ris\des}^{\sf T} \mathbb{E}\{\widehat{\vec{g}}_{\uav\ris} \widehat{\vec{g}}_{\uav\ris}^{\sf T}\} {\vec{g}}_{\ris\des}}{\Vert \vec{g}_{\ris\des} \Vert^2}
    \bigg\} 
    - \mu_{\chi}^2,
\label{eq:var_chi_R}
\end{align}
respectively. 
Then, $|\chi|^2$ follows a SNCCS distribution with scale factor $\frac{\Omega_{\chi}}{2}$, $2 k_\chi$ degrees of freedom, and non-centrality parameter $2\lambda_\chi$, determined by solving the following system of equations
\begin{IEEEeqnarray}{rCl}
&\begin{cases}
    \Omega_\chi (k_\chi + \lambda_\chi) = \sigma_{\chi}^2 + \mu_{\chi}^2, \\
    \Omega_\chi^2 (k_\chi + 2\lambda_\chi) = 2\big( \sigma_{\chi}^2-\frac{1}{2} \big)^2 
        + 4\mu_{\chi}^2 \big( \sigma_{\chi}^2-\frac{1}{2} \big) + \frac{1}{2}, \\
    \Omega_\chi^3 (k_\chi + 3 \lambda_\chi) = 4 \big( \sigma_{\chi}^2-\frac{1}{2} \big)^3 
        + 12\mu_{\chi}^2 \big( \sigma_{\chi}^2-\frac{1}{2} \big)^2 + \frac{1}{2}.
\end{cases}
\label{eq:system_Eq} \\
&\Rightarrow
\begin{cases}
    \Omega_\chi = \rhs_1^{-1} \big( \rhs_2 - {\textstyle\sqrt{\rhs_2^2-\rhs_1 \rhs_3}} \big), \\
    k_\chi = \Omega_\chi^{-1} \big( 2 \rhs_1 - \rhs_2 \Omega_\chi^{-1} \big), \\
    \lambda_\chi = \Omega_\chi^{-1} \big( \rhs_2 \Omega_\chi^{-1} - \rhs_1 \big),
\end{cases}
\end{IEEEeqnarray}
where $\rhs_1$, $\rhs_2$, and $\rhs_3$ are the right-hand side (RHS) of the first, second, and third equality of \eqref{eq:system_Eq}, respectively. 
In addition, $\Vert \vec{g}_{\ris\des} \Vert^2$ also follows a SNCCS distribution with scale factor $\frac{1}{2(\kappa_{\ris\des}+1)}$, $2 N$ degree of freedom, non-centrality parameter $2N\kappa_{\ris\des}$. Hence, following the analysis in \cite{Wells1962}, we obtain \eqref{eq:PDF_A2GSNR_LOS_F}. This completes the proof of Theorem \ref{theo:2}.
\end{IEEEproof}

Regarding Theorem \ref{theo:2}, we have the following comments:

i) Exact closed-form expressions for \eqref{eq:mean_chi_R} and \eqref{eq:var_chi_R} are intractable. 
        Hence, the Jensen bound for a random vector~$\vec{x}$ computed as
\begin{align}
    \mathbb{E}\left\{ \! \frac{\vec{x}}{\Vert\vec{x}\Vert} \! \right\} 
        \succeq \frac{\mathbb{E}\{\vec{x}\}}{\mathbb{E}\{\Vert\vec{x}\Vert\}}
        \succeq \frac{\mathbb{E}\{\vec{x}\}}{\sqrt{\mathbb{E}\{\Vert\vec{x}\Vert^2\}}},
\end{align}
where $\vec{x} \succeq \vec{y}$ indicate that $\vec{x}$ is element-wise greater than or equal to $\vec{y}$, can be adopted to simplify \eqref{eq:mean_chi_R} and \eqref{eq:var_chi_R}.
Hence, 
    ${\mu_{\chi} \ge \frac{\rho_{\uav\ris}}{\bar{\rho}_{\uav\ris}} \sqrt{N} \alpha_\chi}$ 
and ${\sigma^2_{\chi} \ge 1 + \frac{\rho_{\uav\ris}^2}{\bar{\rho}_{\uav\ris}^2} N \beta_\chi}$, where 
$\beta_{\chi} = 1 - \alpha_\chi^2$ and
$\alpha_{\chi} = \frac{\pi}{2} \frac{ L_{1\!/\!2} \left( - \widehat{\kappa}_{\uav\ris} \right) }{\sqrt{\widehat{\kappa}_{\uav\ris}+1}} 
    \frac{ L_{1\!/\!2} \left( -\kappa_{\ris\des} \right) }{\sqrt{\kappa_{\ris\des}+1}}$. 
    
ii) In practices, the number of RIS elements is usually large, e.g., $N = 80$ elements \cite{NguyenTCCN2022} and can be up to $N = 256$ elements \cite{ZhangTVT2023}. Hence, it is reasonable to assume that $N \gg 1$. In this case, due to the central-limit theorem, the distribution of $\snr_\ris$ can be simplified~as
    \begin{align}
    \snr_\ris
    &\sim 
    \mathcal{CN}\Big( 
        \rho_{\uav\ris} \mathbb{E}\{ {\vec{g}}_{\ris\des}^{\sf T} \widehat{\vec{g}}_{\uav\ris} \}, 
        \bar{\rho}_{\uav\ris}^2 \mathbb{E}\{\Vert \vec{g}_{\ris\des} \Vert^2\}
    \Big) \\
    &\sim 
    \mathcal{CN}\Big( 
        \rho_{\uav\ris} N \alpha_\chi, \bar{\rho}_{\uav\ris}^2 N
    \Big).
    \label{eq:scale_snrR}
    \end{align}
    
    Hence, \eqref{eq:PDF_A2GSNR_LOS_F} converges to the SNCCS distribution with scale factor $ \frac{\bar{\rho}_{\uav\ris}^2 \bar{\snr}_\atog N}{2}  \triangleq \frac{\Omega_\ris}{2}$, 2 degrees of freedom, and non-central parameter $2 N \rho_{\uav\ris}^2 \bar{\rho}_{\uav\ris}^{-2} \alpha_\chi^2 \triangleq 2 \lambda_\ris$. 
    
iii) Based on \eqref{eq:cdf_snccs}, the CDF of $\snr_\atog$, denoted as $F_{\snr_\atog}(x)$, is upper-bounded~by 
    \begin{IEEEeqnarray}{rCl}
        F_{\snr_\atog}(x) 
        &\le& p_{\uav\ris}^{\los} F_{\chi_{2}^2}\left( \bar{\rho}_{\uav\ris}^2 \bar{\snr}_\atog N, \rho_{\uav\ris}^2 \bar{\rho}_{\uav\ris}^{-2} \alpha_\chi^2  N; x \right)
        \nonumber\\
        &+&
        p_{\uav\ris}^{\nlos} F_{\chi_{2}^2}\left( \bar{\rho}_{\uav\ris}^2 \bar{\snr}_\atog N, \rho_{\uav\ris}^2 \bar{\rho}_{\uav\ris}^{-2} N; x \right).
        \label{eq:cdf_snrA2G} 
    \end{IEEEeqnarray}

\begin{Conclusion}
Based on \eqref{eq:scale_snrR}, we deduce that, for large number of RIS elements, the cascaded A2G channel can be characterized as a single Rician fading channel.
\end{Conclusion}
\subsection{End-to-End Outage Probability}

The e2e OP of the system, defined as the probability that the minimum/e2e SNR drops below a target threshold ${\snr_{\sf th} = 2^{2 R} - 1}$, where $R$ [bps/Hz] denotes the target spectral efficiency (SE), and is formulated as 
\begin{align}
\textnormal{OP}(\snr_{\sf th})
    &= \Pr\{ \min[\snr_\gtoa, \snr_\atog] < \snr_{\sf th} \} \\
    &=  1 - [1- F_{\snr_\gtoa}(\snr_{\sf th})] [1- F_{\snr_\atog}(\snr_{\sf th})], \\
    &\ge 1 - [1- F_{\snr_\gtoa}^\infty (\snr_{\sf th})] [1- F_{\snr_\atog}^\infty (\snr_{\sf th})],
    \label{eq:op_3}
\end{align}
where $F_{\snr_\gtoa}^\infty (x)$ and $F_{\snr_\atog}^\infty (x)$ are the RHS of \eqref{eq:cdf_G2ASNR_bound} and \eqref{eq:cdf_snrA2G}, respectively. The bound \eqref{eq:op_3} becomes tight when BS transmits with a relatively high power. 
    Moreover, we can find the G2A and the A2G target thresholds, denoted as $\snr_{{\sf th}, \gtoa}$ and $\snr_{{\sf th}, \atog}$, so that $F_{\snr_\gtoa}^\infty (\snr_{{\sf th}, \gtoa}) = L$ and $F_{\snr_\atog}^\infty (\snr_{{\sf th}, \atog}) = L$, which yields the lower bound \
$\textnormal{OP}(\snr_{\sf th}) \ge \textnormal{OP}(\widehat{\snr}_{\sf th})$ with $\widehat{\snr}_{\sf th} = \min\{ \snr_{{\sf th}, \gtoa}, \snr_{{\sf th}, \atog} \}$ is presented in \eqref{eq:e2e_threshold} at the top of this page, where $a_L \!=\! \frac{Q^{-1}(1-L)}{\sqrt{2}}$, $b_N \!=\! \frac{N {\rho}_{\uav\ris}^2}{\bar{\rho}_{\uav\ris}^2}$, and $Q^{-1}(Q(x)) = x$. 

\begin{table*}
\begin{align}
\widehat{\snr}_{\sf th}
    =   \min\left\{
    \left[
        \frac{p^{\los}_{\src\uav} \PL_{\src\uav}^\los}{(\bar{\rho}_{\src\uav}^2)^{M-1}}
        \left( \frac{\kappa_{\src\uav}\bar{\rho}_{\src\uav}^2+1}{\kappa_{\src\uav}+1} \right)^M
        e^{ \frac{M\kappa_{\src\uav}{\rho}_{\src\uav}^2}{\kappa_{\src\uav}\bar{\rho}_{\src\uav}^2+1} }
        + \frac{p^{\nlos}_{\src\uav} \PL_{\src\uav}^\nlos}{(\bar{\rho}_{\src\uav}^2)^{M-1}}
    \right] \frac{L P_\src}{\sigma_\uav^2}, 
    \frac{N P_\uav \bar{\rho}_{\uav\ris}^2}{\sigma_\des^2} 
    \left(
        \frac{p^\los_{\uav\ris} \PL_{\uav\ris}^\los \PL_{\ris\des}}{\left(
            a_L + \alpha_\chi \sqrt{b_N}
        \right)^{-2}}
        + \frac{p^\nlos_{\uav\ris} \PL_{\uav\ris}^\nlos \PL_{\ris\des}}{\left( 
            a_L + \sqrt{b_N}
        \right)^{-2}}
    \right)
\right\}.
\label{eq:e2e_threshold}
\end{align}
\hrulefill
\end{table*}

\begin{Conclusion}
The target spectral efficiency should not exceed $\frac{1}{2}\log_2(1+\widehat{\snr}_{\sf th})$ to maintain the e2e OP at the desired outage level $L$. Here, the factor $\frac{1}{2}$ is due to the fact that two time slots are utilized in the system.
\end{Conclusion}

\subsection{Channel Hardening}

An important phenomenon in RIS-assisted systems is the channel hardening effect, where the channel behaves like a non-fading channel \cite{ChenGLOBECOM2017}. Channel hardening appears in the A2G communication when the following holds \cite{ChenGLOBECOM2017}
\begin{align}
\eta = \frac{\operatorname{Var}\{ |\snr_\ris|^2 \}}{ \mathbb{E}^2\{ |\snr_\ris|^2 \} } 
    =  \frac{ \bar{\rho}_{\uav\ris}^2 + 2 \bar{\rho}_{\uav\ris}^2 N \rho_{\uav\ris}^2 \alpha_\chi^2 }{ [\bar{\rho}_{\uav\ris}^2 + N \rho_{\uav\ris}^2 \alpha_\chi^2 ]^2 }
    =   0.
\end{align}

Therefore, when ${|\rho_{\uav\ris}| = 1}$, e.g., the UAV is hovering at a fixed position or $v = 0$ m/s; and ${N \gg 1}$, channel hardening can occur during the A2G communication. However, even when the UAV is hovering, a variety of perturbations, such as random jittering and random wobbling, hinder the UAV from maintaining still position and prevent channel hardening.

\section{Results and Discussions}

In this section, we present selected results of numerical simulations. Specifically, we consider $f_c = 2$ GHz \cite{3gpp20173gpp}, $\sigma_\src^2 = \sigma_\uav^2 = \sigma^2$, where $\sigma^2 = N_0 + 10\log_{10}B + F$, where $N_0 = -174$ dBm/Hz is the noise spectral density,
$F = 5$ dB is the noise figure, and $B = 10$ MHz is the system bandwidth.
    The sampling period is $T_s = \frac{1}{B}$ MHz to sample both in-phase and quadrature components simultaneously.
The BS, RIS, UAV, and GUE coordinates are $(0, 0, 10)$ $m$, $(150, 0, 25)$ $m$, $(100, 0, 300)$ $m$, and $(200, 0, 1.5)$ $m$, respectively. Environmental coefficients are set to $K_0 = 0$ dB and $K_\pi = 10$~dB.

\begin{figure}
    \centering
    \includegraphics[width=0.9\linewidth]{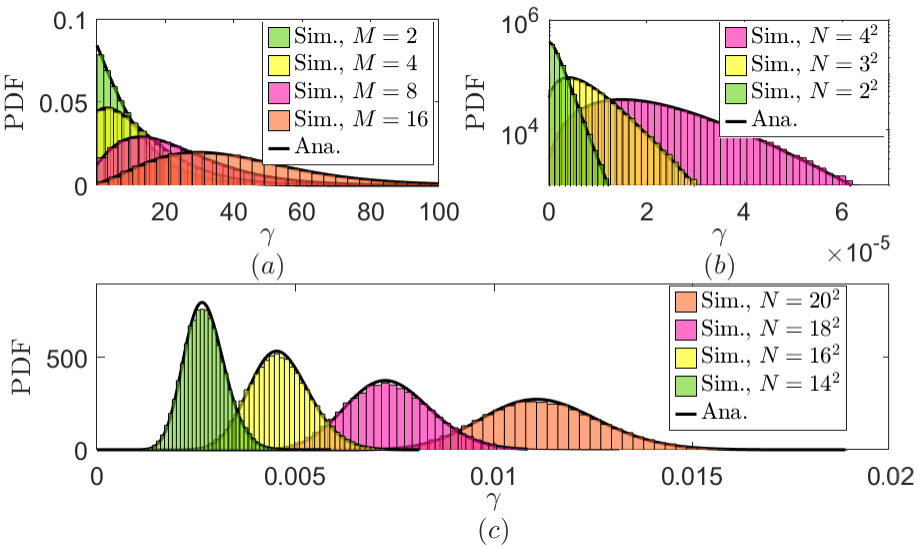}
    \caption{Simulated and analytical PDFs of a) the G2A SNR, b) and c) the A2G SNR, where c) is based on \eqref{eq:scale_snrR} for practical RIS settings. \label{fig1}}
\vspace{-10pt}
\end{figure}

In Fig. \ref{fig1}, where $P_\src = P_\uav = 0$ dBm and $\rho_{\src\uav} = \rho_{\uav\ris} = 0.5$, the analytical and simulated results agree well for different values of $M$ and $N$. To obtain Fig. 1b, we truncate \eqref{eq:PDF_A2GSNR_LOS_F} to 135 terms, where more terms improve the accuracy, but increase the numerical complexity. 
    For large $N$, we utilized \eqref{eq:scale_snrR} to obtain the A2G SNR PDF in Fig. 1c. Across all the figures, we observe a good match between the simulated and derived analytical results. Consequently, the former can be effectively employed to characterize the behavior of the latter.

\begin{figure}
    \centering
    \includegraphics[width=0.9\linewidth]{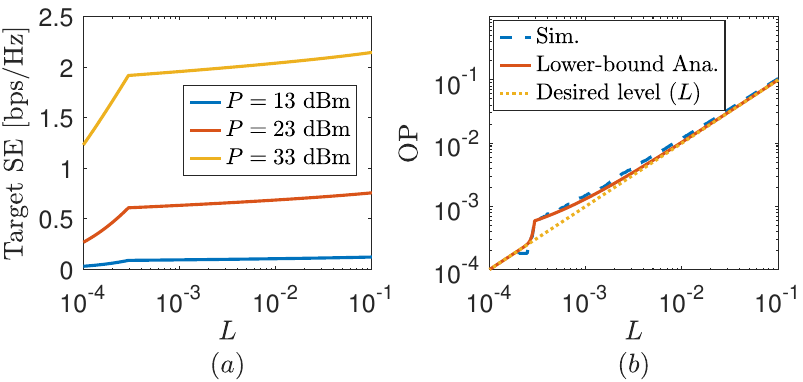}
    \caption{Simulated and derived analytical OP (a) at the target SE $R = \frac{1}{2}\log_2(1+\widehat{\snr}_{\sf th})$ (b) in terms of the required outage level $L$, where $P_\src = P_\uav = P$ [dBm], $M = 4$, and $N = 20^2$.
  \label{fig2}}
\vspace{-15pt}
\end{figure}

Fig. 2a shows an accurate match between the analytical bound in \eqref{eq:op_3} and the simulation results for the OP even for critical desired outage levels, such as $L = 10^{-4}$, which corresponds to $0.01\%$ outage. However, for some intermediate values of $L$, such as $L \in (10^{-3}, 10^{-3.9})$, the OP slightly deviates from the desired outage level. 
    The corresponding maximum target SEs to achieve such OP values are shown in Fig. 2b, which also illustrates how adjusting the transmission power levels can enhance the maximum target SE across varying outage levels.

\begin{figure}[htbp]
  \centering
    \centering
    \subfloat[]{
    \includegraphics[width=0.45\linewidth]{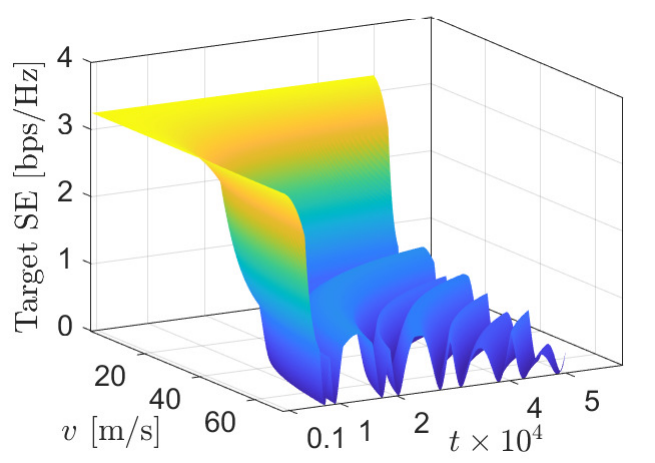}
        \label{fig3a}}
    \subfloat[]{
    \includegraphics[width=0.45\linewidth]{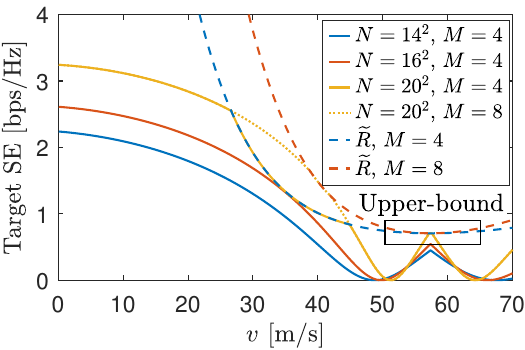}
        \label{fig3b}}
    \caption{Maximum target SE with respect to a) the UAV’s speed ($v$) and sample index ($n$), and b) the UAV’s speed, where $P_\src = P_\uav = 33$ dBm.
  \label{fig3}}
\vspace{-5pt}
\end{figure}

Fig. \ref{fig3} shows the maximum target SE as a function of the UAV speed. We set $R = \frac{1}{2} \log_2(1 + \widehat{\snr}_{\sf th})$ and $L = 10^{-4}$ based on observations in Fig. \ref{fig2}. In Fig. \ref{fig3a}, the target SE decreases with increasing UAV speed or sample index $t$ since $\rho_{\AB} = J_0(2\pi v f_c t T_s c^{-1})$, which influences $\widehat{\snr}_{\sf th}$ and introduces oscillations that cause ripples in the target SE. In Fig. \ref{fig3b}, we use $\widetilde{R} = \frac{1}{2}\log(1+\snr_{{\sf th}, \gtoa})$ as a reference and observe that increasing the number of RIS elements improves the target SE up to a certain upper bound. However, increasing the number of antennas can only enhance the target SE to a certain upper bound, unable to compensate for performance degradation induced by high UAV speeds, e.g., $v \ge 50$ m/s, where a larger RIS or more antennas don't make a difference.

\vspace{-3pt}
\section{Conclusion}
\vspace{-3pt}

This paper studied the statistical characteristics of G2A and RIS-assisted A2G communication under channel aging. Specifically, exact and tractable PDFs for the G2A and A2G SNRs are derived using the analysis for complex Gaussian quadratic RVs. We found that channel hardening can appear in A2G communication when the UAV's speed is low. Moreover, we derived the maximum target SE for the system to operate at various desired outage levels. 
    The results also showed that increasing the number of RIS elements and BS antennas can improve the maximum target SE up to a certain limit; however, high UAV speeds, particularly exceeding $50$ m/s, i.e., $250$ km/h, result in a significant reduction in the target SE that cannot be mitigated by additional RIS elements or additional BS antennas.

\bibliographystyle{IEEEtran}
\bibliography{main}

\end{document}